# Thermo-optic phase shifter based on hydrogen-doped indium oxide microheater

*Weiyu Tong, Erqi Yang, Yu Pang, Haobo Yang, Xin Qian, Ronggui Yang, Bin Hu\*, Jianji Dong\* and Xinliang Zhang*

W. Tong, E. Yang, B. Hu, J. Dong, X. Zhang
Wuhan National Laboratory for Optoelectronics, Huazhong University of Science and Technology, Wuhan 430074, China
E-mail: bin.hu@hust.edu.cn; jjdong@hust.edu.cn
Y. Pang, H. Yang, X. Qian, R. Yang
School of Energy and Power Engineering, Huazhong University of Science and Technology, Wuhan 430074, China

Thermo-optic (TO) phase shifters are very fundamental units in large-scale active silicon photonic integrated circuits (PICs). However, due to the limitation of microheater materials with a trade-off between heating efficiency and absorption loss, designs reported so far typically suffer from slow response time, high power consumption, low yields, and so on. Here, we demonstrate an energy-efficient, fast-response, and low-loss TO phase shifter by introducing hydrogen-doped indium oxide (IHO) films as microheater, and the optimized electron concentration with enhanced mobility endows the IHO high conductivity as well as high near-infrared (NIR) transparency, which allow it to directly contact the silicon waveguide without any insulating layer for efficient tuning and fast response. The TO phase shifter achieves a sub-microsecond response time (970 ns/980 ns) with a π phase shift power consumption of 9.6 mW. And the insertion loss introduced by the IHO microheater is ~ 0.5 dB. The proposed IHO-based microheaters with compatible processing technology illustrate the great potential of such material in the application of large-scale silicon PICs.

## 1. Introduction

Silicon photonic integrated circuits (PICs) trend to lie at the heart of the communications revolution owing to the low power consumption, high speed, large bandwidth and complementary metal-oxide semiconductor (CMOS) compatibility. Recently, emerging applications of PICs such as optical phased arrays (OPAs),[1-3] optical neural networks (ONNs),[4-6] integrated photonic quantum,[7, 8] and so on are also attracting increased interest. The phase shifter is an indispensable fundamental photonic device for its phase tuning capability, and its large-scale integration is in demand in the above various applications. Due to the high thermo-optic (TO) coefficient ($1.84 \times 10^{-4}$ K$^{-1}$) of silicon,[9] phase tuning of silicon waveguide can be achieved easily by heating using a microheater, and such TO tuning as an efficient way in phase shifter has been widely used in large-scale silicon PICs by virtue of the advantages of simple design, easy fabrication, low cost and compactness.[10-13] Unfortunately, the conventional metal-based microheaters for TO tuning require a thick insulating layer (i.e., $SiO_2$ layer) to avoid large absorption loss but would result in high power consumption and slow response speed, which become an obstacle in the application scenarios desiring fast-response TO tuning, such as fast beam scanning of OPAs and fast switching of neurons in ONNs. Recently, designs by placing the metal quite close to the waveguide have been proposed, which can realize low-loss devices based on non-Hermitian system theory, thereby improving the efficiency and speed of TO tuning.[14, 15] However, this approach requires a high-precision fabrication process, having difficulties in large-scale PIC applications. To remove the insulating layer, emerging two-dimension (2D) materials, such as graphene,[16-18] phosphorene,[19, 20] MXenes,[21] and so forth, have been demonstrated to directly overlay waveguides as the microheaters and contribute excellent performances in TO tuning, but the sophisticated preparation and transfer processes of 2D materials also face the difficulties in practical PIC applications due to the low yield. Doped-silicon microheaters can directly generate heat inside the waveguide to achieve high-speed TO tuning,[22-25] but the loss induced by free carriers turns into a challenging issue.

Transparent conductive oxides (TCOs) can be the alternative materials for microheaters, and their mature and reliable mass-production processes provide the practical application in large-scale silicon PICs. So far, the dominant TCO known as tin-doped indium oxide (ITO) has been widely used in optoelectronic devices such as photovoltaic cells, displays, touch screens, etc.[26, 27] Most of TCOs with relatively high free carrier concentration ($n > 5 \times 10^{20}$ cm$^{-3}$) can be transparent in the visible spectral range but always show a strong absorption due to the plasma oscillations in the near-infrared (NIR) waveband,[28] which increases the insertion loss of the phase shifter. Although such an issue can be alleviated by decreasing the doping amount in TCOs for lower $n$, the enhanced NIR transmittance would accompany the decrease of electrical conductivity ($\sigma$) of TCOs which may weaken the heating performance. Fortunately, the conductivity is determined by both carrier concentration and carrier mobility ($\mu$) and can be expressed as $\sigma = en\mu$ ($e$ is the elemental charge). Therefore, enhancing $\mu$ but lowering the $n$ can be an effective way to balance the conductivity and NIR transparency. Nowadays, researchers have demonstrated to replace the tin doped in ITO with other metals for higher $\mu$, such as titanium,[29, 30] tungsten,[31, 32], and molybdenum,[33-35] and prove that the TCO films possess the higher transparency in the NIR waveband and can maintain the comparable σ simultaneously. Nevertheless, the growth temperatures of these TCOs require relatively high (> 300 °C), which exceeds the melting point of photoresists for micro-pattern manufacturing.

In this paper, we propose and experimentally demonstrate a TO phase shifter based on hydrogen-doped indium oxide (IHO) microheater. We fabricate an IHO microheater for phase shifter through room-temperature magnetron sputtering with high-temperature post-annealing treatment. The deposited IHO films possess high $\mu$ and moderate $n$, thus can achieve appropriate sheet resistance and high NIR transparency simultaneously. The compatible fabrication process makes the IHO more suitable for patterning by mature lift-off process and exhibits great potential for on-chip applications. The IHO microheater is designed to directly contact the silicon waveguide without any insulating layer for efficient TO tuning and fast response. The high NIR transparency of IHO ensures low insertion loss of the phase shifter. We measure the TO tuning performance of the proposed phase shifter through the conventional Mach–Zehnder interferometer (MZI) structure. Its fastest response time is 970 ns in rising time and 980 ns in falling time, with a π phase shift power consumption of 9.6 mW, and the insertion loss introduced by the corresponding IHO microheater is about 0.5 dB. Our work illustrates the great potential of IHO to be a novel microheater material for large-scale silicon PICs.

## 2. Results

### 2.1. Chip Design and Fabrication

A phase shifter based on IHO microheater is designed on a silicon-on-insulator (SOI) substrate with a top silicon thickness of 220 nm. **Figure 1**a exhibits the details of the complete fabrication process of the phase shifter. First, we use electron beam lithography (EBL) and inductively coupled plasma (ICP) etching to fabricate the MZI structure. The strip waveguide width is 550 nm and the thickness is 220 nm. Then, the IHO microheaters are directly deposited on the waveguide through magnetron sputtering and the geomtery of the pattern can be well controlled through lift-off process. The entire device is fabricated under standard silicon photonics fabrication processes (no air-trench or undercut process).

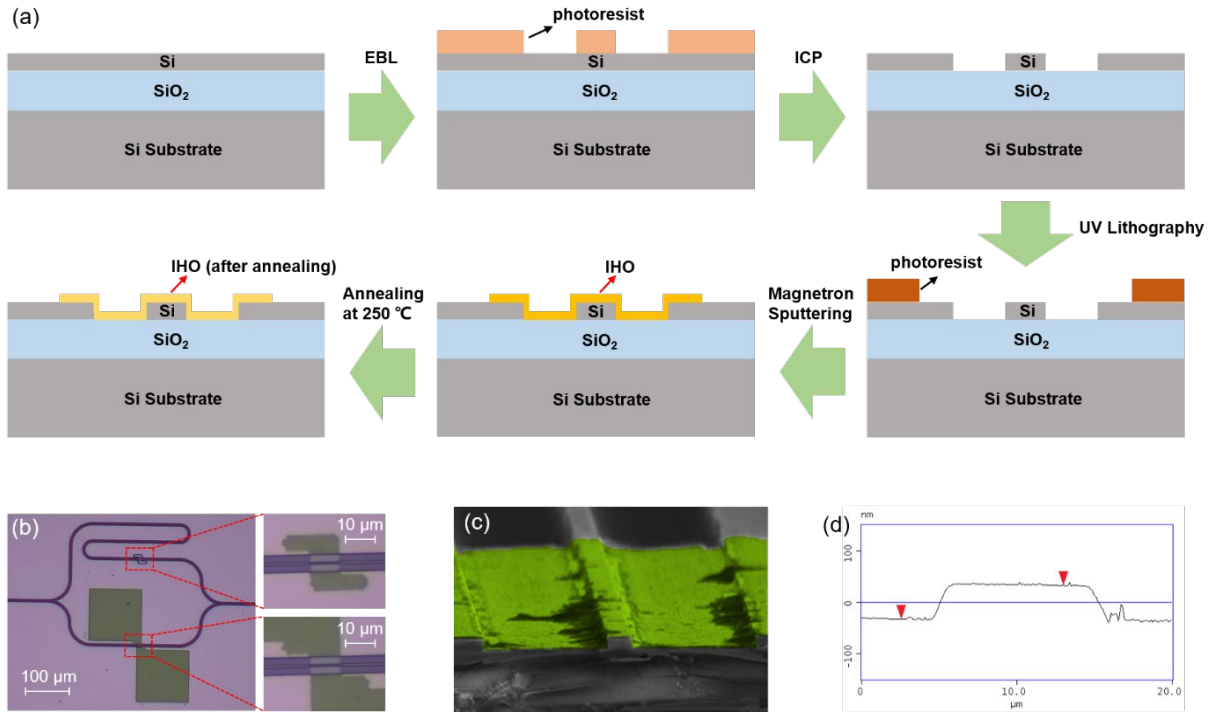

**Figure 1.** a) The schematic of the device fabrication process. b) The microscope image of the fabricated MZI structure. Insets: the zoom-in images of the IHO films. c) The SEM image of the cross-section of the phase shifter. IHO film is false-colored. d)The thickness measurement result of the IHO film.

Figure 1b exhibits a microscope image of the fabricated MZI structure. The insets show the zoom-in images of the IHO films on the waveguides. Both arms of the MZI structure are covered with IHO films to balance the loss, and two additional IHO pads are deposited on both sides of the MZI tuning arm for contact with the probes. The length of the IHO film covering on the waveguide is 10 μm along the light transmission direction. A pair of grating couplers designed for TE polarization are used coupling to the input/output fibers. Figure 1c presents the scanning electron microscope (SEM) image of the cross-section of the phase shifter. The thickness of the IHO film is measured to be 66 nm with an atomic force microscope (AFM), as shown in Figure 1d.

## 2.2. Chip Characterization and Measurements

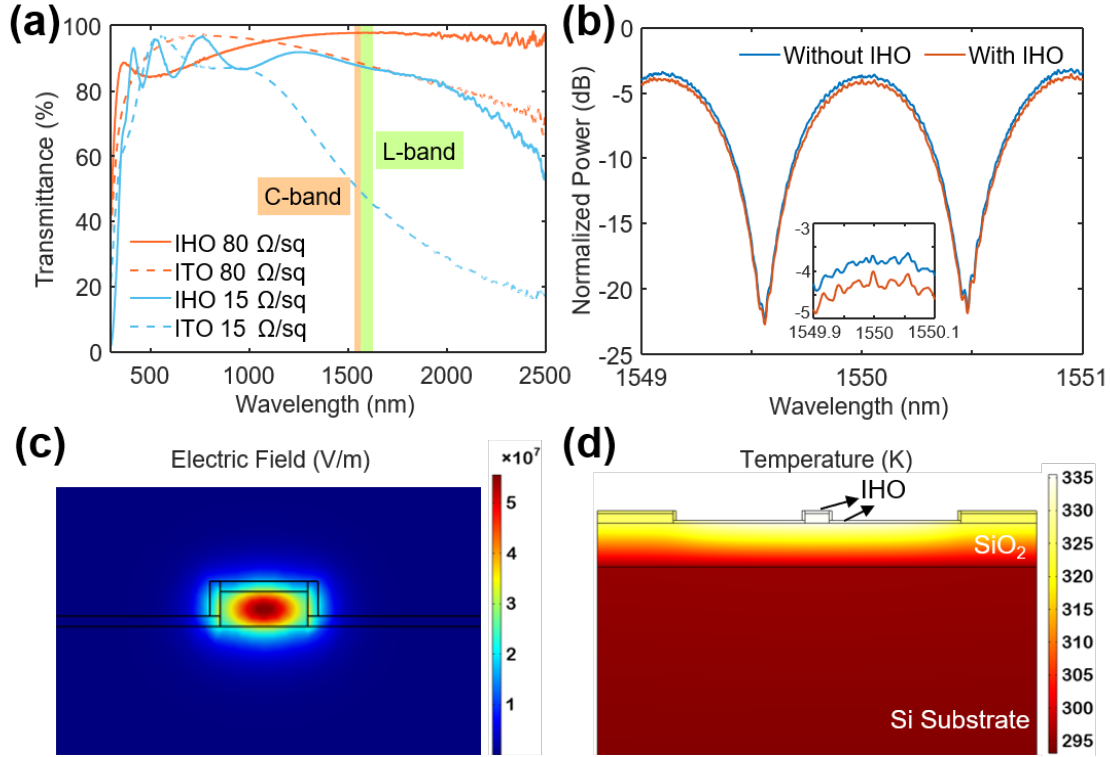

**Figure 2.** a) Comparison of transmittance between IHO and ITO. b) Normalized transmission spectra for the MZI with and without IHO. Insets: the zoom-in images around 1550 nm. c) Simulated optical E field distribution for the fundamental mode at a wavelength of 1550 nm. d) Simulated temperature distribution at a heating power of 9.6 mW.

**Figure 2**a compares the transmittance of the prepared IHO film and commercial ITO film. It can be seen that the IHO possesses significantly higher transmittance than ITO in the NIR waveband under same sheet resistance. In particular, for a 66 nm thick IHO film (corresponding to the solid orange line), it shows a transmittance over 97% in the C+L band (1530 nm - 1565 nm and 1565 nm - 1625 nm). The measured transmission spectra of the fabricated MZI structure with and without IHO microheater are depicted in Figure 2b (orange line and blue line, respectively). The results are normalized to the reference strip waveguide to exclude the coupling loss of the grating couplers. The IHO microheater introduces an excess loss of about 0.5 dB (as shown in the insect) and has little effect on the extinction ratio of the MZI structure. Figure 2c, d present the simulated images of the optical E field distribution and temperature distribution in a cross-section of the phase shifter based on the finite element method (FEM), respectively.

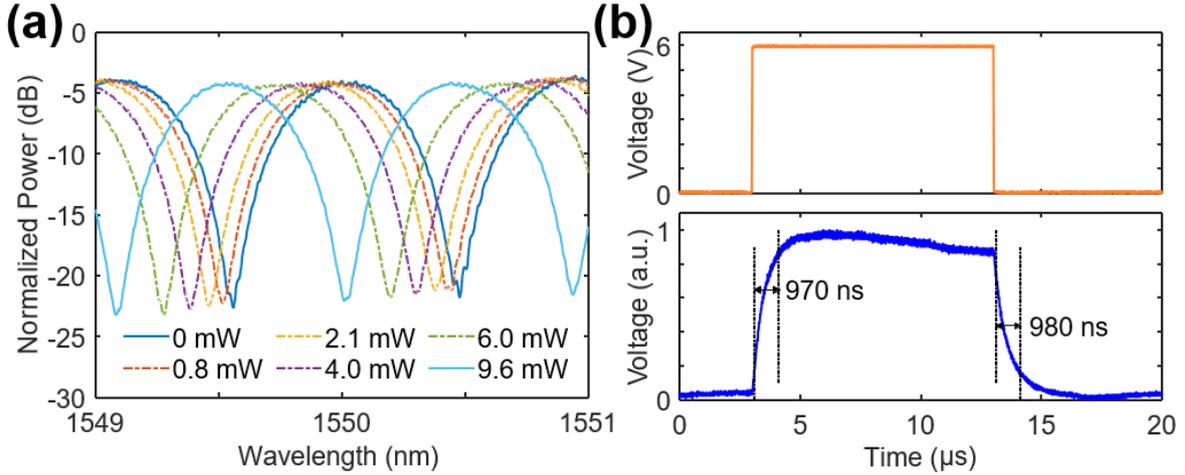

**Figure 3**. a) The transmission spectra at different heating powers. b) The waveform of the driving signal (orange line) and corresponding output signal (blue line).

In order to measure the spectral responses, different driving power are applied to the IHO microheater. The measured spectra are presented in **Figure 3**a. The shift of the interference dip at approximately 1550.48 nm reaches 1/2 free spectral range (FSR) with a tuning power of 9.6 mW. The response time of the device is further characterized by driving the IHO heater with a square waveform electrical signal while the input wavelength is fixed at 1550.47 nm. The frequency of the driving signal is set to 50 kHz with a peak-to-peak voltage ($V_{pp}$) of 6.0 V. The $V_{pp}$ of 6.0 V corresponds to an applied heating power of 9.6 mW. The waveform of the output signal is measured via a photodetector followed by an oscilloscope. The driving signal and corresponding output signal are shown in Figure 3b. The 10–90% rising and falling times are measured to be 970 and 980 ns, respectively.

## 2.3. Performance Comparison

We compared the performance of our proposed phase shifter with the state-of-art TO modulation devices in recent years,[14-16, 18, 20, 22-25, 36-39] as shown in **Figure 4**. The lower π phase shift power consumption ($P_\pi$) of the phase shifter is more beneficial to reduce the overall power consumption of the large-scale PICs, and the shorter response time stands for better system performance. As a comparison, we focus on the cases of MZI structures, and there is no air-trench or undercut process during device fabrication, just as our work. Besides, the response time is estimated by $\tau = 0.35/BW$ if thus parameters were not offered.[24] Compared with other schemes, our work achieves both low power consumption and fast response, second only to 2D material-based microheaters in terms of overall modulation performance. Furthermore, the proposed device structure is simple to fabricate and has great potential for use in large-scale PIC applications.

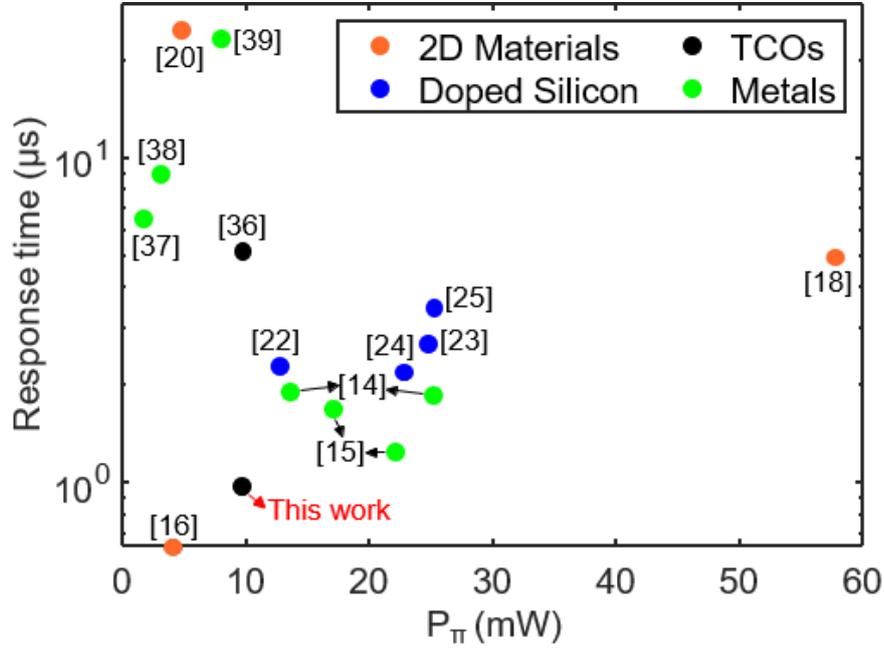

**Figure 4.** Performance comparison of the state-of-art TO modulation devices.

## 3. Conclusion

In conclusion, we propose and experimentally demonstrate a TO phase shifter based on IHO microheater directly overlaying on the silicon waveguide. The phase shifter has an insertion loss as low as about 0.5 dB. Based on the MZI structure, it exhibits a low π phase shift power consumption of 9.6 mW, and a fast response of 970 ns in rising time and 980 ns in falling time. Besides, the IHO film is easy to be patterned on chips. Therefore, our work not only demonstrates an efficient, fast-response, and low-loss TO phase shifter, but also provides a novel design route for optical modulation devices in large-scale PICs.

## 4. Experimental Section

*Materials*: IHO film deposited by magnetron sputtering. Firstly, a high-purity ceramic $In_2O_3$ target (99.99% purity, Zhongnuo Advanced Material Technology Co., Ltd) is used as a sputtering target. The target is put into radio frequency (RF) magnetron sputtering system (Beijing Technol Science Co., Ltd). When the base pressure is less than $10^{-4}$ Pa, the gas that both argon and hydrogen argon mixture (95%Ar and 5%H2) is introduced to the system and various Ar and Ar/H$_2$. After the gas pressure stabilizes at 0.16 Pa, the sputtering power was set at 150 W for 210 seconds at room temperature (300 K). Finally, the as-deposited IHO films were annealed at 250 ℃ for 150 seconds in $N_2$ atmosphere, respectively. All the ITO films was brought by Foshan Shi Yuan Jing Mei Glass Co.,Ltd.

*Device characterization*: We use an amplified spontaneous emission (OS8143) optical source to generate a C-band broadband signal, and a power supply (Rohde&Schwarz HMP4040) to provide a static electrical signal, and finally record the spectral response of the MZI structures by an optical spectrum analyzer (Yokokawa AQ6319). When we measure the dynamic response, the power supply is replaced by an arbitrary waveform generator (RIGOL DG4202), and the input light is emitted from a tunable laser source (ID-Photonics CoBriteDX4). The output signal is measured via a photodetector (Discovery DSC40S) followed by the oscilloscope (RIGOL DG4022). A polarization controller is used to select the TE polarization state of the input light before it is injected into the chip. All the measurements are performed under room temperature in ambient conditions. The specular transmittance of the samples (range from 0.3 μm to 2.5 μm) was characterized by the UV-Vis-NIR spectrophotometer (Shimadzu UV-3600 Plus). .

**Acknowledgements**

This research is supported by the Natural Science Foundation of China (U21A20511, 62274071), the National Key Research and Development Program of China (2022YFB2804200), and the Innovation Project of Optics Valley Laboratory (Grant No. OVL2021BG001). The authors thank the Optoelectronic Micro&Nano Fabrication and Characterizing Facility of Wuhan National Laboratory for Optoelectronics for the support in device fabrication. Weiyu Tong and Erqi Yang contributed equally to this work.

**Conflict of Interest**

The authors declare no conflict of interest.

**Data Availability Statement**

The data that support the findings of this study are available from the corresponding author upon reasonable request.